\documentclass[fleqn,twoside]{article}
\usepackage{espcrc2,epsfig}

\usepackage{graphicx}
\usepackage[figuresright]{rotating}

\newcommand{\AmS}{{\protect\the\textfont2
  A\kern-.1667em\lower.5ex\hbox{M}\kern-.125emS}}

\title{International Lattice Data Grid}

\author{C.T.H. Davies\address[UG]{Department of Physics and Astronomy, 
        University of Glasgow, Glasgow G12 8QQ, UK},
        A.C. Irving\address[UL]{Theoretical Physics Division, Department of Mathematical 
              Sciences, University of Liverpool, Liverpool, L69 3BX, UK},
        R.D. Kenway\address[UE]{School of Physics, University of 
          Edinburgh, The King's Buildings, Edinburgh EH9 3JZ, UK} and
        C.M. Maynard\addressmark[UE]
	for the UKQCD collaboration.}
       
\begin{document}

\begin{abstract}
We propose the co-ordination of lattice QCD grid developments in
different countries to allow transparent exchange of gauge
configurations in future, should participants wish to do so. We describe
briefly UKQCD's XML schema for labelling and cataloguing the data. A
meeting to further develop these ideas will be held in Edinburgh on
19/20 December 2002, and will be available over AccessGrid. 
\end{abstract}

\maketitle

\section{INTRODUCTION}

We envisage that the International Lattice Data Grid (ILDG) would be a
`grid of grids', integrating the emerging national data grids being
developed to manage and exchange lattice QCD data, particularly gauge
configurations~\cite{kenway}. Through co-ordination, we could capitalise
on these grid developments by sharing expertise and tools, and avoiding
duplication of effort. By taking the further step of creating a global
grid infrastructure, we would give added international `backbone' to our
work. 

Since access policies could be determined locally, each collaboration
would still be free to decide which, if any, of its configurations to
share and with whom. Our proposal is complementary to the Lattice Data
Group~\cite{lellouch}, which is compiling a reference set of lattice QCD
results. Clearly, if combined with ILDG, this would add value to the
global grid environment.

The operation of the ILDG would require locally managed, but
interoperable, data grids conforming to a set of standards. Data
formats, labelling, access policies, management and costs would be
handled locally. To achieve this, we will need an international forum to
define the standards for interfacing between the grids and to support
the federation of databases. We will organise a workshop in Edinburgh on
19/20 December 2002 to begin the technical discussions~\cite{workshop}. 

We describe below how an XML schema could be developed for labelling QCD
data. Conformance to XML would allow different schemas to be adopted by
different data grids, with the additional overhead of developing
translation tools to enable each to access the others' data through a
single interface.

\section{XML SCHEMA}

To share resources, either computing or data, one must be able to
describe them in a standard language. To share configurations, one must
have metadata associated with them, containing all the information
about how each configuration was generated and for which physics
parameter values. Most collaborations already have some metadata
describing their configurations, but, to share configurations
automatically, a complete description of the data from the `action' to
the `algorithm' is necessary. The UKQCD collaboration is designing an
XML schema to describe lattice data, which will be extendable, in
principle, to cover any lattice data.

XML is an acronym for eXtensible Markup Language. It is similar to HTML
in that it uses `$<>$' tags. Unlike HTML, it is for structuring data.
XML is license free, platform independent, and well supported. The
commercial web services industry uses XML as standard~\cite{xml}. In
fact, XML is a family of technologies.  One of these is XML schema. This
defines the rules and structure of an XML instance document. It is
designed to allow machines to carry out rules made by people.

XML schemas are designed to evolve, to be updated and extended. XSLT
provides a method of translating one XML instance document conforming
to one schema, into another instance document conforming to different
schema. Thus, each collaboration could have its own schema describing
the form of its own (meta)data. However, this would involve some
duplication of effort, so there would be some benefit if all
collaborations in the ILDG used the same schema. UKQCD is designing a
schema for our QCDgrid which could form the prototype for such a
standard. Our progress can be followed on the UKQCD web
pages~\cite{ukqcd}. Comments are welcome.

\section{METADATA CATALOGUE}

Metadata are mapped into file names by a metadata catalogue. Standard
database tools may be used to search the catalogue and, thereby, to
access a file through a `high-level', physics-based description of it.
We are developing a graphical user interface, initially to access and
manage files, and eventually to build and execute programs using them.
Again, the catalogue and browser could be made available to all the ILDG
collaborations. 

We are keen to encourage the inclusion of both state-of-the-art and
legacy data on the ILDG. Obviously, the former may have considerable
value and access to it might be restricted, at least for an initial
period. The ILDG would facilitate the phased release of data under the
full control of their originator.

Each collaboration would publish the data it wished to make available to
the ILDG on its local node. By browsing the metadata catalogue, someone
wishing to use the data could discover what is available and what usage
conditions apply. Access control can be achieved using software, more
correctly known as middleware, being developed by the HEP experimental
community to meet the challenge of LHC computing, e.g. UKQCD is using
middleware developed by the EU Datagrid Project~\cite{datagrid}, which
is itself built on GLOBUS~\cite{globus}.

\section{CONCLUSIONS}

Compared with a centrally managed respository for gauge configurations,
the ILDG concept offers
\begin{itemize}
\item more flexible data access control (not `all or nothing');
\item the ability to handle different internal data formats easily;
\item distributed management, control and operating costs; 
\item organic growth of both content and functionality; and
\item a basis for greater information and resource sharing in future.
\end{itemize}

We believe that the development of an International Lattice Data Grid
would enhance lattice activity worldwide. The collaboration needed to
define the ILDG standards could avoid duplication of effort and rapidly
promote best practice. If successful as a vehicle for data exchange,
ILDG could expand to support open-source software, algorithm
development, and eventually even shared computers. A global grid
infrastructure would provide a continually growing environment for our
science, encouraging wider participation and facilitating physics
collaborations. As a symbol of international co-operation, it might even
prove useful in making cases for joint machine funding.

\end{document}